\documentclass[preprint2]{aastex63}

\usepackage{CJK}

\shorttitle{Near MMRs Formation}
\shortauthors{Wang et al.}

\begin{document}
\begin{CJK*}{UTF8}{gbsn}

\title{Departure from the Exact Location of Mean Motion Resonances Induced by the Gas Disk in the Systems Observed by Kepler}


\author{Su Wang}
\affiliation{CAS Key Laboratory of Planetary Sciences, Purple Mountain Observatory,\\ Chinese Academy of Sciences, Nanjing 210008, China}

\author{D.N.C. Lin}
\affiliation{Department of Astronomy and Astrophysics, University of California,\\ Santa Cruz, CA 95064, USA}
\affiliation{Institute for Advanced Studies, Tsinghua University, Beijing 100086, China}

\author{Xiaochen Zheng}
\affiliation{Department of Astronomy, Tsinghua University, Beijing 100086, China}

\author{Jianghui Ji}
\affiliation{CAS Key Laboratory of Planetary Sciences, Purple Mountain Observatory,\\ Chinese Academy of Sciences, Nanjing 210008, China}
\affiliation{CAS Center for Excellence in Comparative Planetology, Hefei 230026, China}

\begin{abstract}
The statistical results of transiting planets show that there are
two peaks around 1.5 and 2.0 in the distribution of orbital period
ratios. A large number of planet pairs are found near the exact location of mean motion resonances (MMRs). In this work, we find out that the
depletion and structures of gas disk play crucial roles in driving planet pairs out of exact location of MMRs. Under such
scenario, planet pairs are trapped into exact MMRs during orbital
migration firstly and keep migrating in a same pace. The
eccentricities can be excited.
Due to the existence of gas disk, eccentricities can be damped leading
to the change of orbital period. It will make
planet pairs depart from the exact location of MMRs. With depletion timescales larger than 1 Myr,
near MMRs configurations are formed easily. Planet pairs have
higher possibilities to escape from MMRs with higher disk
aspect ratio. Additionally, with weaker corotation torque,
planet pairs can depart farther from exact location of MMRs. The
final location of the innermost planets in systems are
directly related to the transition radius from optically thick
region to inner optically thin disk. While the transition radius is
smaller than 0.2 AU at the late stage of star evolution process, the
innermost planets can reach around 10 days. Our formation scenario
is a possible mechanism to explain the formation of near MMRs
configuration with the innermost planet farther than 0.1 AU.

\end{abstract}

\keywords{planetary systems: planets and satellites: formation: protoplanetary
disks.}


\section{Introduction} \label{sec:intro}
The Kepler mission and its follow-up program K2 have released over
6000 planetary candidates, including $\sim$ 2700 confirmed planets
and more than 3300 candidates yet to be confirmed \citep{Borucki10,
Bata13, Mazeh13, Fab14, Dr17, Kun20}. Among them, there are hundreds of
confirmed multiple planetary systems. The population of the
discovered planets provides us a good sample to study the dynamics
and formation of planetary systems \citep{MB16, GJ17, MF17, Wang17,
Gong2018, He19, Yang20, Zhang20}. From the statistic results, we find that there are plenty
of planet pairs in the configuration of near MMRs
especially the first order resonances, such as 2:1 and 3:2 MMRs
\citep{Lissauer11,Wang12, lee13, Wang14, Wang17, An20}. However, the peaks
in the distribution of period ratios are shown not in the exact
location of MMRs, but a little departure from them \citep{Wang12,
lee13, Steffen15, Wang17, Wu19, Pan20}. Most of the planet pairs pipe up at the
location larger than the exact MMRs. Near 3:2 MMR, the maximum
fraction of planet pairs can reach $\sim$ 3\% for the period ratio within an interval of 0.025, while the maximum fraction near 2:1 MMRs is about 2\% for the period ratio within an interval of 0.025.
The Transiting Exoplanet Survey Satellite (Tess)
has started operations in July 2018 to detect transiting planets in
85\% of the sky \citep{Ricker15}. Up to now, there are more than 80 planets
have been confirmed and over two thousand planetary candidates
to be confirmed. TOI-125 and TOI-270 which are confirmed by TESS are
multiple planetary systems bearing planets in near MMRs
configurations \citep{Quinn19, Nielsen20}. More systems with similar
configurations are expected to be found by TESS mission to enlarge
the number of planet pairs in near MMRs.

Several mechanisms are proposed to explain the formation of
configuration near exact MMRs. The major scenario of the tidal effect
produces near MMRs for the systems with planets very close to the
central star \citep{PT10, LW12, BM13, Delisle14}. \citet{lee13}
analyzed the systems with planets less than 2$R_\oplus$ in radius
and showed that if they are rocky with $Q/k_2\sim100$, planet pairs
can form near MMRs configuration because of tidal damping. But for
the planets larger than 2$R_\oplus$ or with the distance from the
central star farther than 0.1 AU where the tidal effect is not
strong enough, the formation of near MMRs configuration is not
clear.

\begin{figure*}
\centering
\includegraphics[scale=0.6]{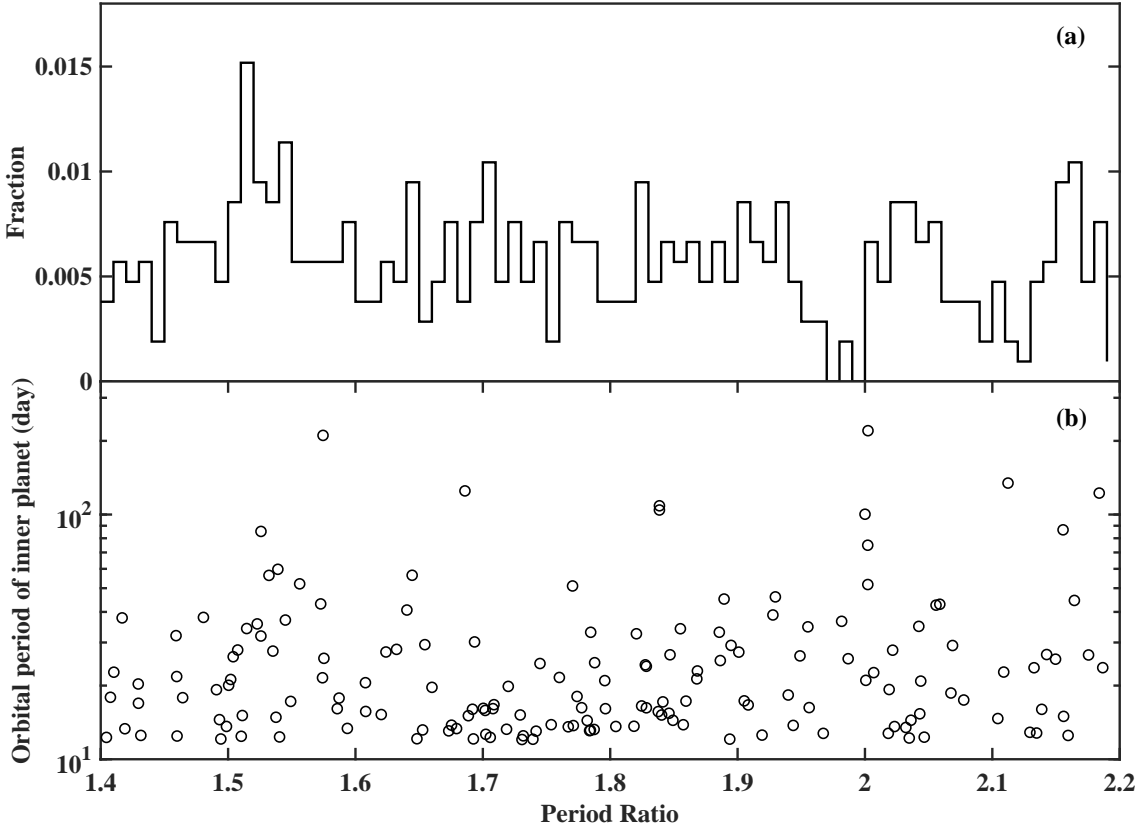}
 \caption{The distribution of period ratio in the system with the innermost planets whose orbital periods are larger than 10 days. Panel (a) shows the fraction of different period ratio in the range of [1.4, 2.2]. Panel (b) displays the distribution of the innermost planets versus the period ratio. The orbital periods of the innermost planets are in a range of 10 - 200 days.
 \label{sta}}
\end{figure*}

The magnetospheric cavity with a one-side torque disk plays an
alternative role in the formation of near MMRs configurations
\citep{Koenigl91, Liu15, Liu17a, Liu20}. They explored the effect of
various planet masses, disk accretion rates, stellar magnetic field
strengths, and depletion timescales of gas disks. The direction of
type I migration in the edge of the cavity is outward. The
configuration of the planet pairs which can be trapped into MMRs at
the very beginning will be rearranged. However, in this scenario, a
special gas disk is required and the final period ratio is related
to the mass ratio of two planets.

A late orbital instability after disk depletion has been suggested
to explain the formation of near MMRs configuration
\citep{iz17, Ogihara18, Lambrechts19}. They assumed that planets
formed  a chain resonance configuration after slow migration with a
timescale of about 1 Myr. The orbital instability during the gas
depletion phase will make close-in super-Earths forming non-resonant
observed configurations. The final period ratio between planets
depends on their masses. But they just investigated the influence of
planet mass on the final stage. The planet mass growth process
\citep{Petrovich13} or interaction with planetesimals \citep{CF15}
can lead to the formation of near MMRs configurations. However, the
depletion timescale and structures of gas disks will affect the
amplitude that departs from the exact MMRs.

Based on the classical migration model, we present a formation
scenario of near MMRs configuration. First of all, planets are
formed at the outer region of the system at which there are enough
material to form terrestrial planets with couple of Earth-mass. Due
to the effect between planets and gas disk surrounded, terrestrial
planets will undergo orbital migration \citep{LP79, KN12}. During
the convergent migration, planet pairs can be captured into exact
MMRs \citep{lee02, PS05, PN08}. If planets migrate to the inner
region of the system, the effect of the tidal raised by the star can
destroy the configuration of MMRs and become near MMRs
configuration. Based on this formation scenario, we investigated the
configuration formation of KOI-152 system with planets in near 4:2:1
MMRs \citep{Wang12}, the formation of near 2:1 and 3:2 MMRs effected
by the properties of star, the speed of type I migration
\citep{Wang14}, the effect caused by the mass accretion process and
the possible outward migration \citep{Wang17}, and the existence of
giant planets in planetary systems \citep{Sun17,Pan20}.

We analyse the data with the innermost planets in a system farther
than 0.1 AU. The results are shown in Figure \ref{sta} with 1054
planet pairs. We mainly focus on the region near 2:1 and 3:2 MMRs in
the range of [1.4 2.2] in the distribution of period ratio between
planet and its adjacent inner planet. Panel (a) shows the fraction
distribution. We find that near the exact location of 2:1 MMRs, where a gap is shown
between the period ratio of 1.97 and 2.0. Few cases are found in
this region. A peak appears at the period ratio larger than 2.0
especially the region between 2.0 and 2.06. The fraction decreases
from 0.01 to 0.001 in the region of [2.06 2.13]. Similar tendency is
given near the exact location of 3:2 MMRs. Few planet pairs are found between the period
ratio of [1.4 1.5]. A peak exists at the period ratio larger than
1.5 and smaller than 1.55. The fraction decreases from 0.015 to
0.004 in the region of [1.55 1.62]. Another peak appears at about
1.69 which is near 5:3 MMR. The fraction between 1.63 and 1.68
remain stable, is consistently lower than 0.09. Panel (b) of Figure
\ref{sta} shows the distribution of the orbital period of innermost
planets versus the period ratio between two adjacent planets. Most
of the innermost planets are in the range of [10 100] days, where
seven planet pairs with the period of inner planet larger than 100
days. In this work, our primary goal is to investigate the formation
scenario of systems with semi-major axis of the inner planet larger
than 0.1 AU, along with planet pairs in near exact MMRs configuration.

We suppose that planet pairs can be captured into exact MMRs first during
the orbital migration process which induced by the gas disk. Then
planet pair will migrate in a same pace and eccentricities of
planets can be excited after they are in MMRs. The depletion
timescale of gas disk is estimated to be million years
\citep{hai01}. Here, we assume the gas disk surrounding with the
star can survive from $10^5$ to $10^7$ yr \citep{WC17}. If the
eccentricity damping effect is still strong enough, eccentricities
which have been excited in MMR will be damped. Due to the
eccentricity damping, the semi-major axis will change a little.
Thus, planet pairs will depart from the exact location of MMRs. In this formation scenario, the
depletion and structures of gas disks are possible essential factors
that influence the final configurations.

The work is structured as follows, in Section 2, we describe the
models including disk models and numerical methods used in our
simulations. The main results of simulations containing some typical
cases and the statistical results are summarized in Section 3.
Section 4 shows the conclusions and discussions.

\section{Models}
\subsection{Disk Models}
The surface density profile of gas  disk  $\Sigma_{\rm g}$  based on
the empirical minimum-mass solar nebular model (MMSN; Hayashi 1981)
at a stellar distance $r$ is described as
\begin{eqnarray}
\Sigma_{\rm g}=f_{\rm g}\Sigma_0(\frac{r}{1AU})^{-k} {\rm
exp}(\frac{-t}{\tau_{\rm dep}})\rm ~g~cm^{-2}~~~~~~~~~~~~~~~~~~
\nonumber\\
=1700 f_{\rm g} (\frac{r}{1AU})^{-1} {\rm exp}(\frac{-t}{\tau_{\rm
dep}})\rm ~g~cm^{-2}~~~~~~~~~~~~~~~
\label{gas}
\end{eqnarray}
where $\tau_{\rm dep}$ is the gas depletion timescale and $f_{\rm
g}$ is the enhancement factor of the MMSN, \emph{k}=1 is the
power-law index of the gas density we adopted in this work, and
\emph{t} means the time.

\subsection{Type I migration and gas damping}
We adopt the prescription for type I migration as described in
\citet{Paa10, Paa11} where the total torque can be expressed in
terms of
\begin{eqnarray}
\Gamma_{tot}=f_{tot}\Gamma_0/\gamma~~
\nonumber\\
f_{tot}=f_{LB}+f_{CR}
\end{eqnarray}
where $f_{tot}$, $f_{LB}$ and $f_{CR}$ are the coefficients for the
total, Linblad, and corotation torques, respectively, $\gamma=1.4$
is the adiabatic index, and
\begin{equation}
\Gamma_0=(q/h)^2\Sigma_pr_p^4\Omega_p^2,
\end{equation}
where $\Sigma_p$ and $\Omega_p$ are the disk surface density and
angular frequency at the location of the planet $r_p$, $q=M_p/M_*$,
and $h=H/r$ is the disk's aspect ratio at $r_p$. The magnitude of
$f_{tot}$ is a function of $s\equiv \partial ln\Sigma/\partial r$,
$\beta \equiv \partial ln T/\partial r$, $q$, $\nu$ (viscosity), and
$\xi$ (thermal diffusivity). Herein, we assume an expression to approximate the $f_{tot}$ as

\begin{eqnarray}
f_{tot}=f_{nsc}Q(q)+f_{LB}~~~~~~~~~~~~~~~~~~~~~\\
\nonumber\\
f_{nsc}=coef\times(1-\frac{2(\frac{r}{r_t})^2}{1+(\frac{r}{r_t})^2})~~~~~~~~~~~~\\
\nonumber\\
Q(q)=\frac{(\frac{q}{q_1})^{2}(\frac{q}{q_2})^{-2}}{2(\frac{q}{q_1})^{2}(\frac{q}{q_2})^{-2}+(\frac{q}{q_1})^{2}+(\frac{q}{q_2})^{-2}}\\
\nonumber\\
f_{LB}=1-\frac{2(\frac{r}{r_m})^4}{1+(\frac{r}{r_m})^4},~~~~~~~~~~~~~~~~~~~~~
\end{eqnarray}
where $r_t$ is transition radius between the outer optically thick
region and the inner optically thin disk, it decreases with the mass
accretion rate, gas density and time. $r_m$ is magnetosphere radius. $f_{nsc}$ is the coefficient of the fully non saturated component of the corotation torque, the saturation parameter $Q(q)$ represents a range of $q_1<q<q_2$ over which the corotation torque is not saturated.
$coef$ is coefficient of $f_{nsc}$. With different $coef$, the
strength of the corotation torque will change.

\begin{figure*}
\begin{center}
\includegraphics[scale=0.65]{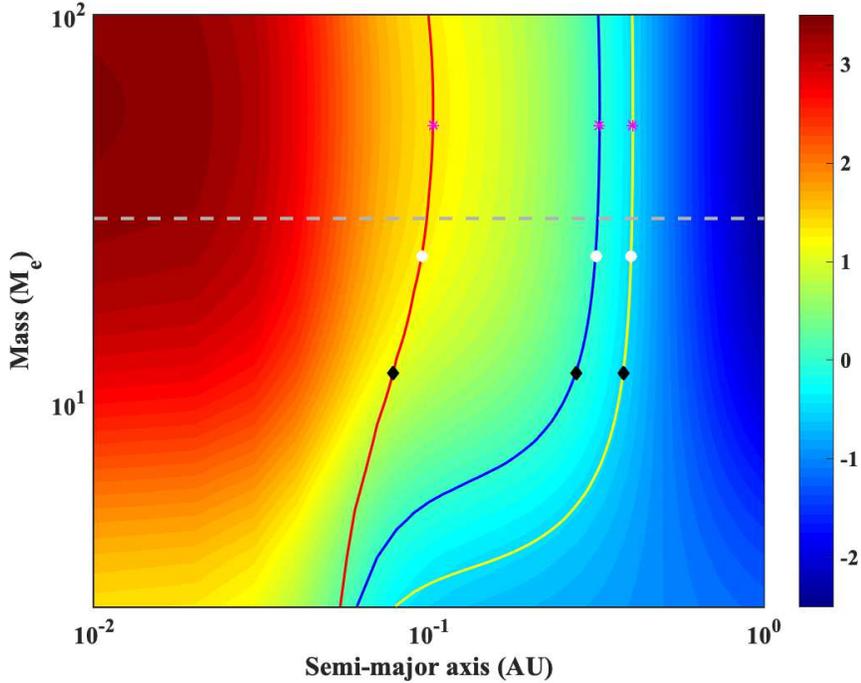}
 \caption{The contour of total torque $f_{tot}$ varying with the planetary mass and semi-major axis. The colorbar index denotes the values of the  total torque, where red color represents the positive torque while the blue stands for the negative torque. If the semi-major axis is less than $r_{boundary}$, planets will undergo outward migration. The red, blue and yellow lines represent the location of $r_{boundary}$ with respect to $coef=2$, $coef=5$ and $coef=10$, respectively. The black diamonds show the exact locations of $r_{boundary}$ for the planets with 12 $M_\oplus$ which is the mass of the inner planet in our simulations, while the white dots display the exact locations  of $r_{boundary}$ for the planets with 24 $M_\oplus$ which is the mass of the outer planet in our simulations. The magenta stars represent the furthest location on each boundary line, it locates at 0.1 AU for $coef=2$, 0.32 AU for $coef=5$, and 0.41 AU for $coef=10$, respectively.
 \label{torq}}
 \end{center}
\end{figure*}

If $f_{tot}$ is negative, planet will undergo inward migration with
the negative torque. On the contrary, if $f_{tot}$ is positive, the
planet will suffer from outward migration with the positive torque.
Figure \ref{torq} shows the value of $f_{tot}$ changing with the
mass of planets and the distance away from the central star. If the planet is massive enough ($m>M_{crit}$) \citep{idalin08}, 
\begin{equation}
M_{\rm crit}\simeq 30(\frac{\alpha}{10^{-3}})(\frac{a}{1AU})^{1/2}(\frac{M_*}{M_\odot})M_\oplus,
\end{equation}
a gap will form around it. The torque on planet is non-linear due to gap opening and planet will undergo type II migration other than type I migration. $M_{crit}$ is about 30 $M_\oplus$ for the system with solar-like star. In Figure \ref{torq}, the grey dash line shows the location of 30 $M_\oplus$. The torque we estimated in this work is suitable for the region below the grey line. In
Figure \ref{torq}, red color means that $f_{tot}$ is positive, while
blue color indicates that $f_{tot}$ is negative. As observed from
Figure \ref{torq}, we can conclude that planets will experience
outward migration within 0.25 AU when the planet mass is about 10
$M_\oplus$. Here we define this boundary between inward and outward
migration as $r_{boundary}$, which $r_{boundary}$ changes with the
transition radius $r_t$. Meanwhile, $r_{boundary}$ is related to the
planet mass. In Figure \ref{torq}, we assume $r_t=0.5$ AU, $coef=5$,
$q_1=3\times 10^{-5}$ $M_\odot$ and $q_2=10^{-3}$ $M_\odot$.
$r_{boundary}$ appears to remain between 0.2 to 0.32 AU when the
planetary mass ranges from 10 to 300 $M_\oplus$ as shown in the blue
line. The timescale of type I migration is
\begin{equation}
\tau_a=\frac{m_p \sqrt{GM_*r}}{2\Gamma_{total}}=\frac{m_p\gamma
\sqrt{GM_*r}}{2f_a(\frac{q}{h})^2\Sigma_gr^4\Omega^2}=\frac{\tau_0}{f_ah^2}.
\label{tau_a}
\end{equation}
Herein, $f_a=f_{tot}$. The timescale of gas damping is
\begin{equation}
\tau_e=h^2f_a\tau_a=\frac{m_ph^2\gamma
\sqrt{GM_*r}}{2(\frac{q}{h})^2\Sigma_gr^4\Omega^2}=\tau_0.
\label{tau_e}
\end{equation}
where $M_*$ represents the mass of central star and we choose
$M_*=1~M_\odot$ in this work. $\Omega$ is the Kepler angular
velocity, and $G$ is the gravitational constant.

The force of eccentricity damping is expressed as
\begin{equation}
\textbf{F}_{damp}=-\frac{(\textbf{v}\cdot
\textbf{r})\textbf{r}}{r^2\tau_e}
\end{equation}
The force of type I migration is
\begin{equation}
\textbf{F}_{migI}=\frac{\Gamma_{tot}}{m_pr}
\end{equation}

The acceleration of the planetary embryos with a mass $m_i$ is
described as

\begin{eqnarray}
\frac{d}{dt}\textbf{V}_i =
 -\frac{G(M_*+m_i
)}{{r_i}^2}\left(\frac{\textbf{r}_i}{r_i}\right) ~~~~~~~~~~~~~~~
\nonumber\\
+\sum _{j\neq i}^N
Gm_j \left[\frac{(\textbf{r}_j-\textbf{r}_i
)}{|\textbf{r}_j-\textbf{r}_i|^3}- \frac{\textbf{r}_j}{r_j^3}\right]
\nonumber\\
+\textbf{F}_{\rm damp}+\textbf{F}_{\rm migI},~~~~~~~~~~~~~~~
\label{eqf}
\end{eqnarray}

In this work, we integrate equation (\ref{eqf}) to explore the
dynamical evolution of planets in the system using the time
symmetric integrator Hermit scheme \citep{Aarseth}. In our numerical
simulations, all planets are initially assumed to occupy coplanar
and circular orbits. The mean anomaly and the argument of pericenter
are generated between $0^{0}$ to $360^{0}$ randomly. Each case is
integrated to 10 Myr.

\section {Numerical Simulation Results}

In our model, we consider the planetary systems are composed of two
planets with a couple of Earth-mass and a solar-mass central star.
Herein we perform extensive numerical simulations to investigate the
dynamical evolution of planet pairs in the system, where we take
into account the combined parameters of a wide variety of density of
gas disk $\Sigma_g$ which reflects the change of $f_g$, the
depletion timescale of gas disk $\tau_{\rm dep}$, the coefficient of
the torque $coef$, the transition radius $r_t$, the disk aspect
ratio $h$. $p_{10}$ and $p_{20}$ are the initial orbital periods of
two planets, the subscripts 1 and 2 represent the values of inner
planet $P_1$ and outer planet $P_2$, $m_1$ and $m_2$ respectively
denote the planetary masses. This allows us to extensively explore
the dynamical nature of planet pairs at specific MMRs using the
above-mentioned parameters.

\begin{figure*}
\begin{center}
\includegraphics[scale=0.45]{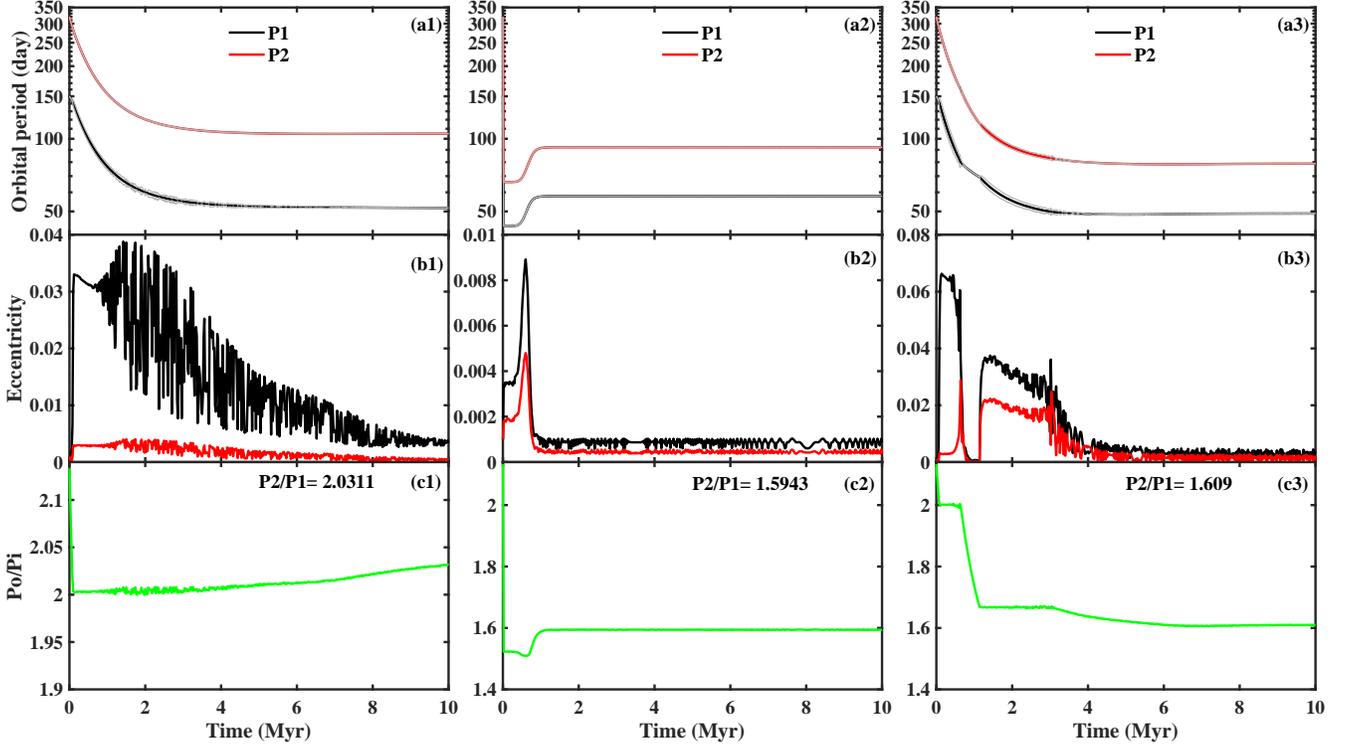}
 \caption{The evolution of Case 1-3. Three typical runs show that two planets finally deviates from exact MMR. Panel (a1-3), (b1-3), and (c1-3) display the evolution of the orbital periods, semi-major axes, eccentricities, and the period ratio, respectively. The black line and red lines, respectively represent the evolutions of orbital periods for the inner planet ($P_1$) and the outer planet ($P_2$). In Panel (a), the grey lines are associated with the evolution of the pericenter and apocenter, respectively.
 \label{d21}}
 \end{center}
\end{figure*}

We entirely carry out 288 runs in two groups with the combination
parameters as above-mentioned. The details of parameters we adopted
in each group are shown in Table \ref{groups}. In the following, we
show five typical runs (Cases 1-5) revealing the evolution scenario
that two planets deviate from exact MMRs. Table \ref{cases}
summarizes the adopted initials for three runs, where $p_{1e}$ and
$p_{2e}$ represent the resultant orbital periods of the pairs,
respectively.

\begin{table*}
\centering \caption{The parameters used in two groups. $p_{10}$ and
$p_{20}$ are the initial orbital period of two planets in the
systems. $m_1$ and $m_2$ are the masses of planets. The main
difference between group 1 and 2 is the transition radius $r_t$.
 \label{tb1}}
\begin{tabular*}{18cm}{@{\extracolsep{\fill}}cccccccccc}
\hline
 Group& $\tau_{dep}$& $r_t$&$coef$& $h$&$f_g$&$p_{10}$&$p_{20}$&$m_{1}$&$m_{2}$\\
 &Myr &(AU) &&&&(day)&(day)&$(M_\oplus)$  &($M_\oplus)$\\
\hline
1&0.1, 0.5, 1, 3 &0.5 &2, 5, 10 &0.02, 0.05, 0.1, 0.05$r^{1/4}$&1/50, 1/10, 1&150&320&12&24\\
2&0.1, 0.5, 1, 3 &0.2 &2, 5, 10 & 0.02, 0.05, 0.1, 0.05$r^{1/4}$&1/50, 1/10, 1&150&320&12&24\\
\hline
\label{groups}
\end{tabular*}
\end{table*}

\begin{table*}
\centering \caption{The parameters used in four cases. $p_{10}$ and
$p_{20}$ are the initial orbital period of two planets in the
systems. $m_1$ and $m_2$ are the masses of planets. $p_{1e}$ and
$p_{2e}$ represent each of the final orbital periods, respectively.
$f_g$ is enhance factor of the standard gas disk.
 \label{tb1}}
\begin{tabular*}{16cm}{@{\extracolsep{\fill}}cccccccccccc}
\hline
 Case& $\tau_{dep}$& $r_t$&$coef$& $h$&$f_g$&$P_{10}$&$P_{20}$&$m_{1}$&$m_{2}$&$P_{1e}$&$P_{2e}$\\
 &Myr &(AU) &&&&(day)&(day)&$(M_\oplus)$  &($M_\oplus)$&(day)&(day)\\
\hline
1&1 &0.5 &2 & 0.05&1/50&150&320&12&24&51.69&104.99\\
2&0.1 &0.5 &5 & 0.02&1/1&150&320&12&24&57.85&92.23\\
3&1 &0.5 &2 & 0.1&1/10&150&320&12&24&49.03&78.89\\
4&1 &0.5 &5 & 0.05&1/500&300&540&24&12&245.70&495.58\\
5&1 &0.5 &5 & 0.05&1/500&300&640&12&24&264.57&529.61\\
\hline
\label{cases}
\end{tabular*}
\end{table*}

\subsection{Escape process from the exact location of MMR}
After two planets are trapped in exact location of MMR during orbital migration, they will keep migrating in a same pace. The eccentricities of them can be excited in this process.
Due to the existence of gas disk, eccentricities can be damped leading
to the change of orbital period at the same time. The change of angular momentum caused by the eccentricity damping force is zero, therefore, we can get that
\begin{equation}
\Delta a\propto ae\Delta e/(1-e^2) \simeq ae^2,
\label{delt}
\end{equation}
where $\Delta a$ is the change of semi-major axis caused by the eccentricity damping force. 
If eccentricities of two planets are excited to comparable values, the change of semi-major axes $\Delta a$ of them are mainly proportional to $a$. The outer planet will change larger distance than the inner one due to the eccentricity damping process. If the eccentricity of inner planet is excited to value which is much larger than the outer one, the change of period ratio is related to $ae^2$ which depends on the relative space between two planets and the amplitude of eccentricity after they are excited.

\subsection{Case 1: Departure from the Exact Location of 2:1 MMR}
Panel (a1), (b1) and (c1) of Figure \ref{d21} show a typical run for
two planets departing from 2:1 MMR. Here the depletion timescale of
gas disk is assumed to be 1 Myr and the gas density is fiftieth of
the standard MMSN model. According to description of surface density of gas disk in equation (\ref{gas}), $\Sigma_g$ will decrease to $\Sigma_0/e$ at 1 Myr at the same location, but the gas is still exist and eccentricity damping effect is working in the next few million years. As noted from Figure \ref{d21}, two planets
undergo an inward type I migration before the inner planet reaches
$\sim$ 50 days at about 2 Myr. The planet pair is captured into exact 2:1
MMR quickly. Planets will remain at the exact location of 2:1 MMR if
there is no eccentricity damping exist. According to the estimation of equilibrium eccentricity in MMR (equation (A27) of \citet{PS05}), the amplitude of eccentricity is affected by the mass ratio, period ratio and the ratio between timescale of eccentricity damping and orbital migration \citep{MD99,NP02,PS05,Liu15}. Eccentricity tends to be excited to higher value for less massive inner planet. In case 1, in the first two million
years, the eccentricity of the inner planet can be excited to be
approximately 0.04, whereas that of the outer planet is not
radically stirred up but performs a slight fluctuation about 0.005.
In Panel (b1) of Figure \ref{d21}, we observe that the eccentricity
of $P_1$ falls down to 0.01 gradually due to the eccentricity
damping from 2 to 4 Myr. In this case, the excited eccentricity of
$P_1$ when two planets are trapped into exact 2:1 MMR is much higher than
that of $P_2$, $e_1/e_2\sim 10$, while $a_1/a_2$ almost keeps in
0.63 for 2:1 MMR. Thus, according to equation (\ref{delt}), $\Delta a_1$ is larger than $\Delta a_2$
with the damping of eccentricities, the space between two planets
become larger and larger. Panel (c1) of Figure \ref{d21} displays
the evolution of the ratio of orbital period  between two planets
$P_2/P_1$, which was started at about 2.13 (not in 2:1 MMR), then
temporarily to be 2.0 (captured into exact 2:1 MMR with resonant angles librating in very small amplitude), and a final value of
2.0317 (leaving out of exact location of 2:1 MMR with resonant angles librating in large amplitude).  Clearly, this gives a dynamical
portrait for elucidating the resonant evolution history for two
planets in the system. As given in Table \ref{cases}, we have their
resultant orbital periods of 51.69 and 104.99 days, respectively,
being suggestive of that two planets do slightly deviate from exact
2:1 MMR.

From the Kepler data, we can present several systems which hold
planets near the exact location of 2:1 MMRs. Such as the systems Kepler-328 and Keper-56.
Kepler-328 \citep{Xie14} horbors two planets confirmed in 34.92 and
71.31 days, the period ratio is 2.0421. Kepler-56 \citep{Steffen13}
owns two planets in 10.50 and 21.41 days, with an orbital period ratio of  2.0390. These two systems may be formed through the similar
scenario as Case 1 describes.

\subsection{Case 2: Departure from the Exact Location of 3:2 MMR}
The parameters used in this case are shown in the second line of Table
\ref{cases}. Comparing with Case 1, the density of the gas disk is
larger in this case, using the standard MMSN model, and the
coefficient of $f_{nsc}$ is 5 which is larger than that of Case 1.
Therefore, the  boundary $r_{boundary}$ is much larger in Case 2
than in Case 1. The red, blue, and yellow lines in Figure \ref{torq}
show the distribution of $r_{boundary}$ with $coef=2$, $coef=5$ and
$coef=10$, respectively. At the boundary location, the coefficient
of total torque is 0, and when planet approaches the boundary, the
speed of orbital migration will be reduced. We note that the
boundary line declines as $coef$ decreases. For the inner planet
which holds 12 $M_\oplus$, $r_{boundary}$ moves from $\sim$ 0.08 AU
(corresponding to the orbital period of 8.5 days) to 0.27 AU
(corresponding to the orbital period of 51 days) when we change
$coef$ from 2 to 5 as shown in the black diamonds in Figure
\ref{torq}. $r_{boundary}$ is the largest when $coef=10$, the
boundary locates at about 0.38 AU (corresponding to the orbital
period of 85 days). For the outer planet with mass in 24 $M_\oplus$,
$r_{boundary}$ moves from $\sim$ 0.1 AU ( 10 days) with $coef=2$ to
0.32 AU  (66 days) with $coef=5$ as shown in white dots in Figure
\ref{torq}. Therefore, it seems to be possible for inner planet
whose semi-major axis is smaller than 0.27 AU to migrate outward in
Case 2.

The evolution process of Case 2 is shown in panel (a2), (b2), and
(c2) of Figure \ref{d21}. In such a case, $f_g=1$, the gas density
is fifty times of the value given in Case 1, suggesting a much
faster inward orbital migration. As a result, this contributes to
the major reason inducing two planet passed through 2:1 MMR and
trapped into 3:2 MMR. When two planets are in the location near 2:1
MMR, the torque on the planet is still strong enough, thus they will
continue migrating inward until they stop at the region near 43.8
and 66.6 days, respectively. Different with Case 1, the inner planet
run through the location of $r_{boundary}$ to the inner region which
may cause outward orbital migration. Within 0.9 Myr, they are kept
in the exact location of 3:2 MMR with small amplitude of resonant angles as shown in panel (c2) of Figure
\ref{d21}. The eccentricities of inner planet can be excited to be
about 0.01. Subsequently, decrease of semi-major axis is caused due
to eccentricity damping. According to equation (\ref{delt}), the
variation of semi-major axis of $P_1$ caused by eccentricity damping
is larger than that of $P_2$. Considering outward orbital migration
and the effect of eccentricity damping simultaneously, two planets
migrate outward in 0.1 Myr, and the outer planet migrate outward for
a longer distance than the inner planet. Finally, two planets
eventually arrive at 57.85 and 92.23 days respectively with a ratio
of orbital period 1.5943 which deviates from the exact location of 3:2 MMR with large amplitude resonant angles. This indicates a scenario of planet pair
departs from the exact location of 3:2 MMR owing to eccentricity damping and outward
orbital migration.

Kepler-276 and Kepler-279 \citep{Rowe14} are the systems with
planets near the exact location of 3:2 MMRs. There are three planets confirmed in both
systems. Planets locate at 14.13, 31.88, and 48.65 days in
Kepler-276. The outer two planets are in near 3:2 MMRs. The period
ratio of them is about 1.5260. The configuration of Kepler-279 is
similar to Kepler-276. Three planets are in 12.31, 35.74, and 54.42
days, respectively. The period ratio of the outer two planets is
about 1.5227. The formation process showed in case 2 may be a
possible explanation about the formation of these two systems.

\subsection{Case 3: Departure from the Exact Location of 5:3 MMR}

Compared with Case 1, we adopt a disk aspect ratio $h=0.1$ and the
gas density herein $f_g=0.1$, being tenth of the standard MMSN model
in Case 3. According to Equation (\ref{tau_a}) and (\ref{tau_e}),
the migration timescale $\tau_a$ in Case 3 is 0.8 times of that in
Case 1 and the eccentricity damping timescale $\tau_e$ in Case 3 is
3.2 times of that in Case 1. Thus, the eccentricity damping process
can exist for longer time in Case 3.

Similarly with Case 1, we remain $coef=2$ in the calculation.
$r_{boundary}$ locates at about 0.08 AU for inner planet and 0.095
AU for outer planet. Panel (a3), (b3), and (c3) of Figure \ref{d21}
show the evolution of two planets. From the evolution of period
ratio in Panel (c3), we note that two planets are trapped into 2:1
MMR at the very beginning. With faster speed of orbital migration,
two planets break through 2:1 MMR and perform continuous  migrating
into the inner region, and they will be further captured into 5:3
MMR with resonant angles librating at very small amplitude at $\sim$ 1 Myr. Since the inner planet is less massive, the eccentricity of it should be excited to larger value than the outer one . However, the ratio of eccentricity between the outer and inner planet $e_2/e_1$ increases with the decrease of $a_2/a_1$ \citep{NP02, PS05}. Therefore, in this case, eccentricities of both planets
can be stirred up to comparable values, about 0.03.  Under such
circumstance, both planets smoothly migrate when the eccentricities
damp down. And at this time, two planets locate at the outer region
of $r_{boundary}$. Therefore, they will migrate inward. Due to the
similar $e$ and $\Delta e$, $\Delta a$ is proportion to $a$
according to equation (\ref{delt}). Thus, the outer planet will move
inward with longer distance than the inner planet. The simulation
results are consistent with the theoretical estimation. Consequently, the
final ratio of their orbital periods is 1.6099 as shown in panel
(c3) of Figure \ref{d21}, implying that two planets move a little
out of 5:3 MMR. In the end, two planets halt migration and remain at
the orbits of 49.03, and 78.89 days, respectively. Planet pair is out of MMR with cicular resonant angles.

The system Kepler-197 \citep{Rowe14}, which holds four planets, has
planet pair in the configuration similar to that shown in Case 3.
Planet d and e locate at about 15.68 and 25.21 day which make the
period ratio of them equal to 1.6078. The system Kepler-154 has
planet pair in near 5:3 MMRs. There are five planets in the system
\citep{Rowe14, Morton16}. The third planet, Kepler-154 d, and the
fourth planet, Kepler-154 b, are in the period of 20.55 and 33.04
day, respectively. Their period ratio is about 1.6078, which is
the same as the planet pair in Kepler-276. These systems may share
the similar formation scenario with Case 3.

\subsection{Case 4: Departure from the Exact Location of 2:1 MMR with Divergent Migration}

The systems in Group 1 and 2 are initially set to have two planets and the mass of
outer planet is larger than the inner one. Base on the estimation of
isolation mass of planet embryos \citep{idalin04}, $m$ is
proportional to $a^{3/4}$. The outer planet tends to contain more
material than the inner one. According to the statistics on the
Kepler candidates \citep{Wang17}, there is opportunity for stars to
bear a larger innermost planet in a system. Herein, we run a case on
the system with higher mass inner planet to test the effect of the
eccentricity damping. 

\begin{figure*}
\begin{center}
\includegraphics[scale=0.5]{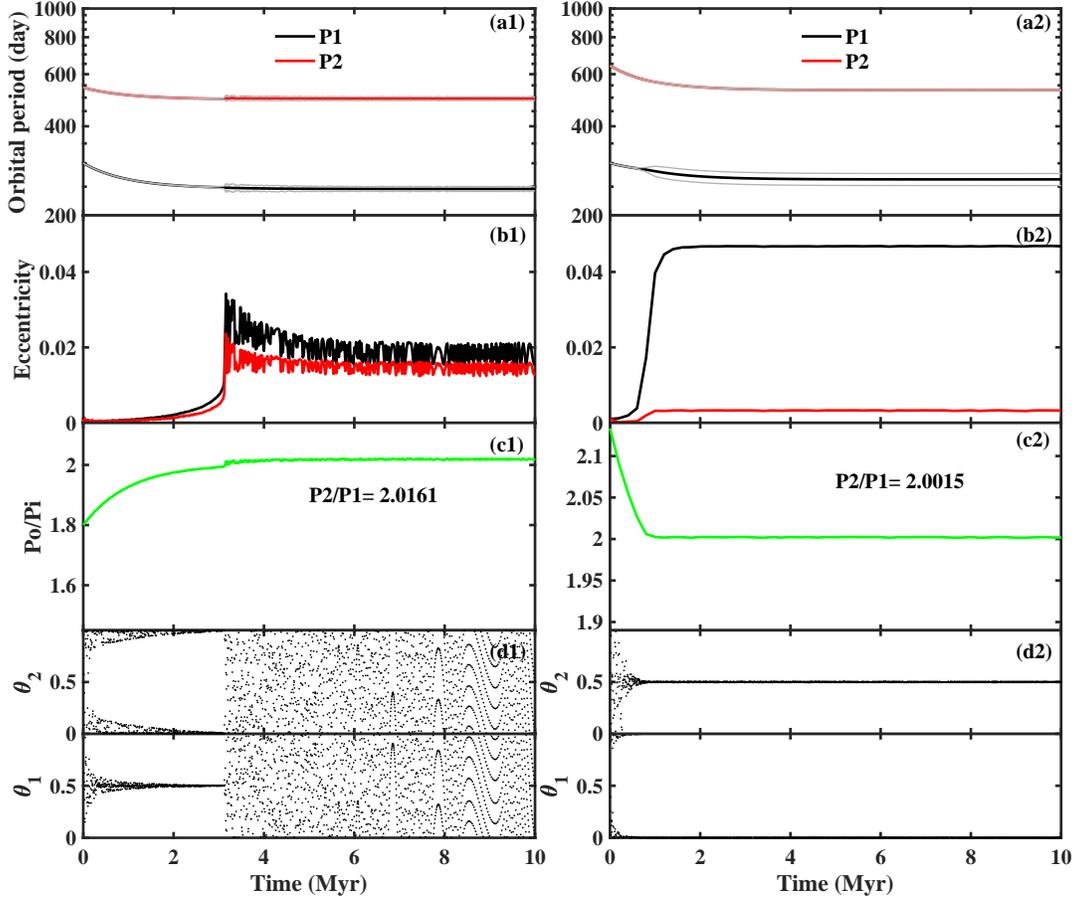}
 \caption{The evolution of case 4 and 5. Panel (a1) and (a2) show the evolution of orbital periods, panel (b1) and (b2) display the evolution of eccentricity, panel (c1) and (c2) represent the evolution of period ratio between two planets of each case, panel (d1) and (d2) show the evolution of resonant angles, where $\theta_1=(2\lambda_2-\lambda_1-\varpi_1)/2\pi$, $\theta_2=(2\lambda_2-\lambda_1-\varpi_2)/2\pi$, $\lambda$ and $\varpi$ are the mean longitude and the longitude of pericentre of each planet, respectively.
 \label{outone}}
 \end{center}
\end{figure*}

In this case, the mass of the inner planet is
24 $M_\oplus$, and the mass of the outer one is 12 $M_\oplus$. The results are shown in Figure \ref{outone}. The transition
radius is 0.5 AU, the scale heigh of disk is 0.05, $coef=5$,
$f_g=0.002$ and the depletion timescale of the gas disk is 1 Myr.
Details of the initial settings in this case are shown in Table \ref{cases} as case 4. Due to the low gas density of the disk,
the migration process is very gentle. The initial periods of two
planets are 300 and 540 day, respectively. The period ratio of them
is smaller than 2.0 initially. After about 3 Myr divergent
migration, the period ratio of planet pair will increase from 1.8 to
2.0, two planets are trapped into 2:1 MMRs, the resonant angle librate with small amplitude as shown in panel (d1) of figure \ref{outone}. Since the inner planet is more massive, the eccentricity of it should be excited to smaller value than the outer one . However, $e_2/e_1$ is also related to $g=(m_1a_2)/(m_2a_1)$\citep{NP02, PS05}. With increase of $g$, $e_2/e_1$ decreases. $g$ in case 4 is larger than that in case 1. Thus, eccentricities of two planets become more comparable in case 4 than in case 1. The eccentricities of
them can be excited to be 0.03 as shown in Panel (c) and the
eccentricity of the inner planet is a little higher than that of the
outer one. Then the eccentricities can be damped due to the gas disk
and finally the period ratio of planet pair can reach 2.0175, which is
slightly larger than 2.0. The resonant angles become circular, planet pair is out of MMR as shown in panel (d1) of figure \ref{outone}.  Comparably, we run case 5 with similar initial conditions as in case 4, but switch the order of the two planets. The initial conditions are shown in Table \ref{cases} and the evolution process is shown in figure \ref{outone}. Two planets are captured into 2:1 MMR at about 1 Myr. Eccentricity of the inner planet can be excited to about 0.05. The 2:1 MMR is more robust through convergent migration than through divergent migration, thus eccentricity damping effect with lower gas density ($f_g=1/500$) cannot make the planet pair in case 5 leave the exact location of MMR, but can destroy the MMR configuration in case 4.

According to our formation scenario, systems
which hold a larger inner planet than the outer one can form the
configuration of near MMRs. Noticeably, two planets locate at about
0.8 AU and 1.2 AU at the end of the simulation. This case can
explain the systems near MMRs configuration with the innermost
planet quite far away from the central star. Additionally, if
planets locate in the inner region of the system initially where
planets will undergo outward migration, they can also get out of the
exact MMR to be near MMR. The process is similar to that as shown in
Case 1-3. The difference is that with outward orbital migration, the
innermost planet can reach the location farther away from the
central star. Considering all the cases in this work, with the
eccentricity damping of the gas disk, planet pairs can get out of
the exact MMRs to be the configuration of near MMRs.

\subsection{Statistical Results}

\begin{figure*}
\begin{center}
\includegraphics[scale=0.65]{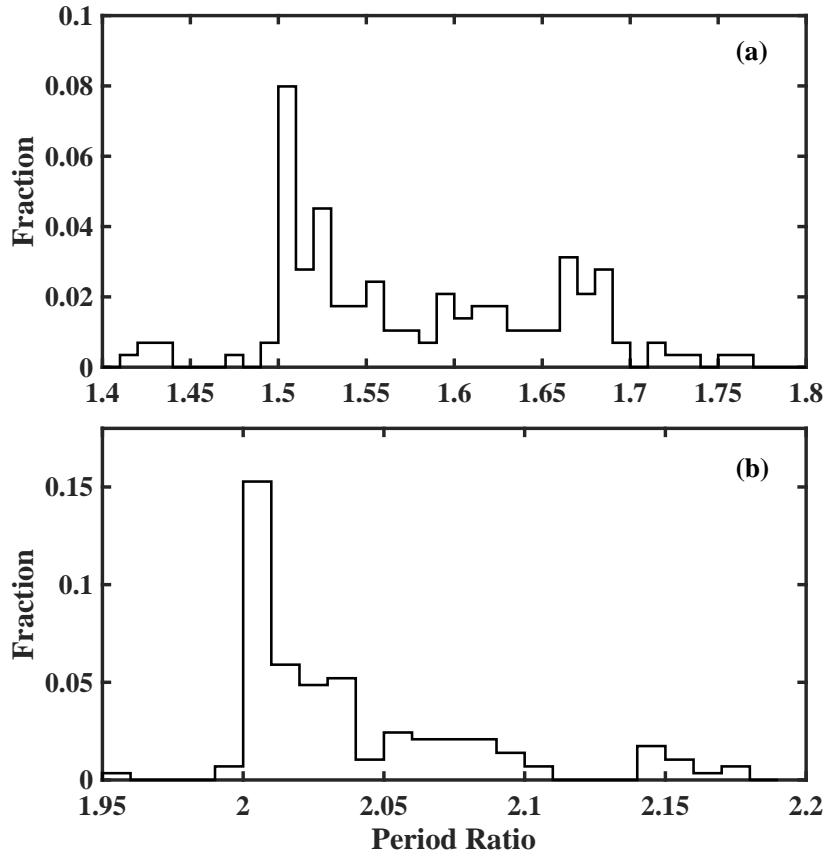}
 \caption{Distribution of period ratio between two planets at the end of the simulations. Panel (a) shows the planet pairs which are in near 3:2 and 5:3 MMRs. Panel (b) displays the distribution of planets locating near 2:1 MMR.
 \label{finalp}}
 \end{center}
\end{figure*}

We entirely perform a set of 288 simulations by considering a
combination of gas density, coefficient, timescales of gas disk, the
disk aspect ratio, and transition radius. The detailed initial
settings are given in Table \ref{groups}. Figure \ref{finalp} shows
the distribution of the period ratio between two planets at the end
of the simulations. Panel (a) displays the fraction of period ratio
in the range of [1.4 1.8], which means that two planets are in near 3:2
and 5:3 MMRs. Panel (b) exhibits the fraction of planet pairs in
near 2:1 MMR, where the period ratio is in the range of [1.95 2.2]. From
Figure \ref{finalp}, we find that two peaks appear to be near 3:2 and 2:1
MMRs obviously. Particularly, there is a deficit of planet pairs
with period ratio smaller than exact 2:1 and 3:2 MMRs, an excess of
pairs with period ratio larger than 2:1 and 3:2 MMRs. The results
are consistent with the observational distribution shown in Figure
\ref{sta} (Lissauer et al. 2011). Another small peak appears at the
position where the period ratio larger than 1.66 slightly. The
fraction of period ratio between 1.5 and 1.66 experiences a decrease
from 8\% to 1\%, while the fraction of period ratio larger than 2.0
see a decrease from 15\% to 1\%. These trends can be seen from the
observation data in Figure \ref{sta}. From table \ref{groups}, we
know that the initial periods of two planets are 150 and 320 day.
The period ratio between them is about 2.13, which is larger than
2.0. Therefore, planet pairs are trapped into 2:1 MMR first and more
easier to be kept in near 2:1 MMR, leading to
the absence of data between 1.77 and 2.0 which is inconsistent with
the observation result. With the same reason, few systems in our
simulations hold planets with period ratio lager than 2.1. In summary, the distribution of period ratio which obtained through the
formation scenario we proposed is consistent with the observation
results.

\begin{figure*}
\begin{center}
\includegraphics[scale=0.65]{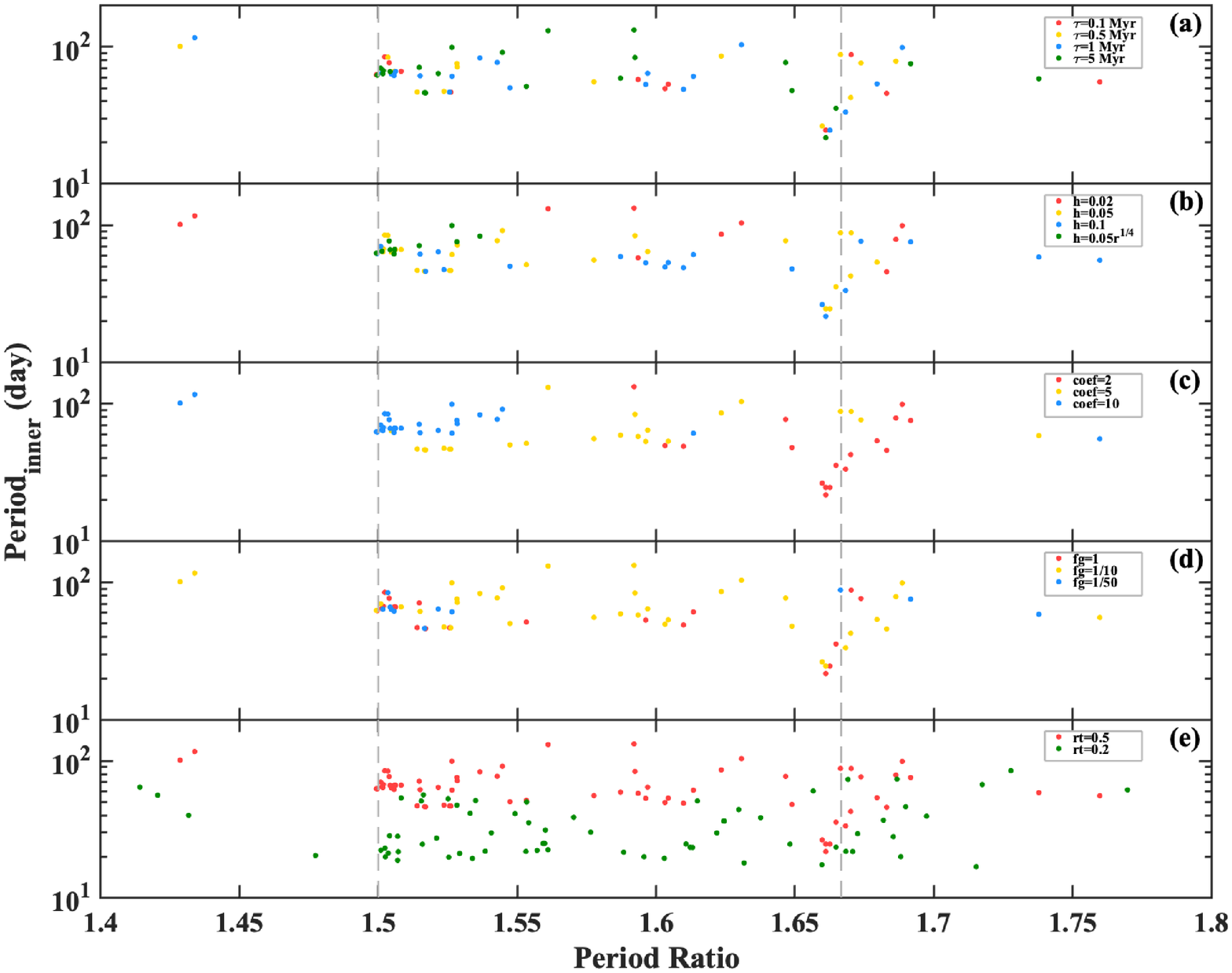}
 \caption{Distribution of planet pairs in the space of period ratio between 1.4 and 1.8 versus the period of inner planet. Panel (a)-(d) show the distribution of planet pairs changed with the depletion timescale of gas disk $\tau$, disk aspect ratio $h$, coefficient of $f_{nsc}$ $coef$, enhance factor of gas density $f_g$, and transition radius $r_t$, respectively.
 \label{f15}}
 \end{center}
\end{figure*}

\begin{figure*}
\begin{center}
\includegraphics[scale=0.65]{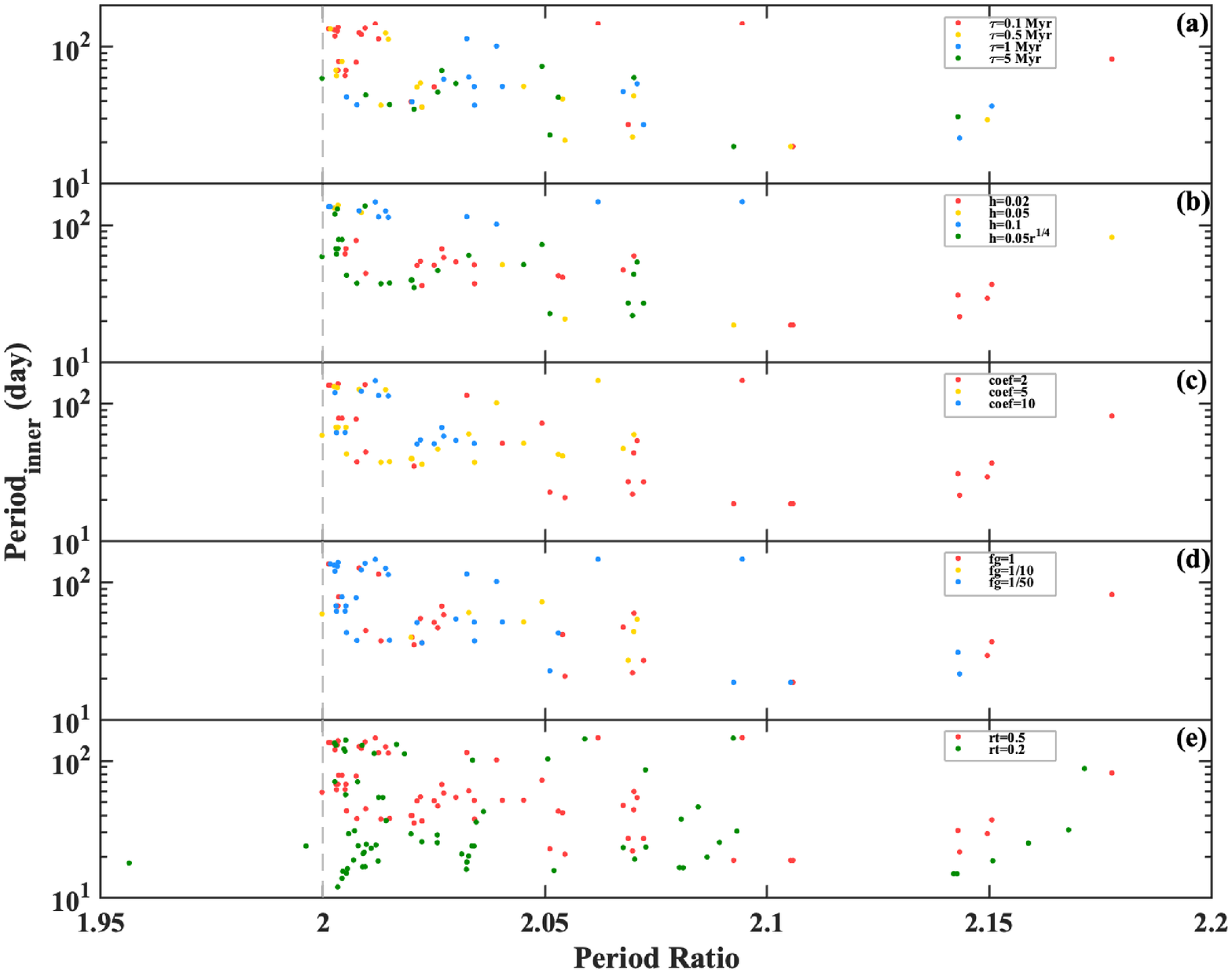}
 \caption{Distribution of planet pairs in the space of period ratio between 1.95 and 2.2 versus the period of inner planet. Panel (a)-(e) show the distribution of planet pairs changed with the depletion timescale of gas disk $\tau$, disk aspect ratio $h$, coefficient of $f_{nsc}$ $coef$, enhance factor of gas density $f_g$, and transition radius $r_t$, respectively.
 \label{f20}}
 \end{center}
\end{figure*}

Figure \ref{f15} and \ref{f20} show the distribution of period ratio
varies with different parameters of the gas disk. Figure \ref{f15}
shows the distribution of period ratio between 1.4 and 1.8, while
Figure \ref{f20} displays the period ratio distribution from 1.95 to
2.2. The period ratios of few cases are less than 1.4, we choose 1.4
as the minimum period ratio as shown in these figures. Panel (a)-(d)
of two figures exhibit the results obtained through 144 runs
in Group 1 with $r_t=0.5$. The overall trend with data from
Group 2 with different disk parameters is similar to that in Group
1. There are 288 dots in Panel (e) with all results from Group 1 and
2, which reveal the major difference with different $r_t$.

From Figure \ref{f15} and \ref{f20}, we find that a great number of
planet pairs are involved in the position very close to the exact
locations of 2:1 and 3:2 MMRs, which is consistent with the results
shown in Figure \ref{finalp}. From Panel (a) of Figure \ref{f15} and
\ref{f20}, we get that planet pairs are more easily to be in near
3:2 MMR rather than 2:1 MMR with longer disk depletion timescale
especially for the system with $\tau=5$ Myr labeled in green dots,
while more planet pairs are in the configuration near 2:1 MMR with
$\tau \le 0.5$ Myr. The possibilities that planet pairs in near 3:2
and 2:1 MMRs are almost even. The results demonstrate that planet
pairs have long time to get rid of the 2:1 MMR which is the first
low order MMR they meet. With longer disk depletion timescale, $\tau
> 1$ Myr the period ratio of planet pair is wide spreading in the
whole region from 1.5 to 2.15, most of them in the range of [1.5
1.7] and a small part of them distribute at the period ratio of [2.0
2.15]. With $\tau \leq 0.5$ Myr, most planet pairs contribute to
pipe-up of period ratio near 2.0, especially from 2.0 to 2.02. 

Based on the formation scenario, planet pair is trapped into MMR quickly through orbital migration process and eccentricities of them will be excited, the eccentricity damping effect is still strong enough to work on planets making them depart from the exact location of MMR with longer disk depletion timescale. Therefore, longer disk depletion timescale is more helpful in planet pairs departure from the exact location of MMRs.

In all cases, planets locate less than 1 AU, thus
$h=0.05r^{1/4}$ is the most smallest one among the $h$ value we
chosen. Through Panel (b), we show that, with higher $h$ value, longer
timescale of eccentricity damping, planet pairs have more
opportunities to escape to the configuration that depart farther
from the exact location of MMRs.

Moreover, we notice that most of these cases with $coef=10$ which
are marked by blue dots in Panel (c) are in the region with period
ratio less than 1.55 (or 2.04 for the planet pair in near 2:1 MMR).
With the decrease of $coef$, planet pairs leave far from their
exact region of MMRs.

From equation (\ref{tau_a}), we learn that $f_g$ is related to
the density of gas disk which determine the speed of orbital
migration directly. With higher gas density, larger $f_g$, planet
pairs are more easily to break through 2:1 to be captured 3:2 MMR
(Wang et al. 2014). Panel (d) shows the results consistent with the
estimation. Most of the cases with $f_g=1/50$ labeled in blue dots
distribute near the 2:1 MMR.

Based on our analysis of the torque acted on the planets, with the
decrease of the transition radius $r_t$, the orbital periods of
inner planets will decline. Combined with 144 cases with $r_t=0.2$,
we can conclude that most of the inner planets lying in the range
[20, 250] days with $r_t=0.5$, while the periods of inner planets
can be much closer to 10 days with $r_t=0.2$ as shown in Panel (e)
of Figure \ref{f15} and \ref{f20}. With the evolution of the star,
the transition radius will migrate inward. Therefore, the systems
with inner planets lying near 10 days may form at the late stage of
the star.

Additionally, from Figure \ref{f15} and \ref{f20}, we can obtain
that the resonant offset from 2.0 can extend to almost 2.2 and most
planet pairs pipes up in the region $2.0 \leq P_2/P_1\leq2.1$. The
period ratios of planet pairs which near 1.5 and 1.667 can extend in the range of [1.3, 1.7]. Only few cases with larger scale
height can be maintained in the region between 1.8 and 2.0.

\section{Conclusions and Discussions}

In this work, we mainly investigate the formation of planetary
systems in the configuration of near MMRs, especially for the
systems with the semi-major axis of innermost planets larger than
0.1 AU. Considering the eccentricity damping effect induced by the
gas disks, we entirely perform 288 runs of simulations under the
condition of a wide variety of depletion timescales of gas disk, the
disk aspect ratio, the enhance factor of the gas density, the
coefficient of the co-rotation torque, and transition radius of the
disk. From the simulations, we conclude that with proper
eccentricity damping effect, planet pairs can deviate from perfect
MMRs and tune in near MMRs, being indicative of that our proposed
scenario is likely to throw light on the distribution nature of the
observed systems. Here we summarize the major results as follows:

\begin{enumerate}
\item
According to our formation scenario, planet pairs can be trapped
into 2:1, 3:2 or 5:3 MMRs through orbital migration with
different migration speed.  Specifically, a large number
of simulations linked to 5:3 MMRs are produced with $h=0.05$ or
$h=0.1$. For those systems harboring the innermost planets with a
distance larger than 0.1 AU, the formation of near MMRs can be
elucidated by the existence of proper depletion timescales of the
gas disks.  Due to the eccentricity damping induced by the gas disk,
planet pairs can move out of the exact location of MMRs and be involved in near MMRs
configurations. From our simulations, we show that the depletion of
the gas disk can make the distribution of the period ratio
comparable to the statistics of the observation.
\item
With the depletion timescales larger than 1 Myr, we find that near
MMRs configurations are easily to form. After planet pair is trapped
into MMR, eccentricity damping is still strong enough to make planets
move adequate distance to depart from exact MMR. Eccentricity
damping plays a crucial role in leading to the deviation from the
exact MMRs for planet pairs. Additionally, with the decrease of
$coef$, which represents the strength of corotation torque, planet
pairs can depart farther from the exact MMRs. Moreover, planet pairs
have higher possibilities to escape from the configuration of MMRs
with higher disk aspect ratio which means longer eccentricity
damping timescales.
\item
The final orbital periods of the innermost planets are directly
related to the transition radius. If the innermost planets ranges in
[50, 200] days, the transition radius of the disk is probably at
about 0.5 AU. In the case of the transition radius near 0.2 AU, the
innermost planets can arrive at the orbital period 10 days. This suggests
that the system with innermost planet closer to the central star
holds stable planets at the later stage of star evolution process.
\item
If the period ratios of planet pairs locate in the range of [1.8,
2.0], planets may be formed in the system with the gas disk of larger
disk aspect ratio.
\end{enumerate}

Tidal effect plays a significant part in giving rise to the
departure of exact MMRs for the system with the extreme close-in
innermost planets the central star \citep{lee13}. From our work, we
can conclude that after the planet pairs are captured into MMRs due to
the migration process, planet pairs can depart from the exact MMRs
locations because of the eccentricity damping caused by the gas disk
for the systems with the innermost planets farther than 0.1 AU in
which the tidal effect of the star is not strong enough to affect
the final configuration of the systems. Through our simulations, we
can explain the distribution of the period ratio between the
adjacent planets observed by the Kepler Mission. The timescale of
the depletion of gas disk and disk aspect ratio are related to how
far the planet pair departs from the exact MMRs and the location of
the transition radius is connected with the final location of the
innermost planets. The scenario can be also applied to explain the formation of planetary systems observed by  TESS which may find a number of planet pairs in near MMRs \citep{Quinn19, Nielsen20}.

\section*{Acknowledgments}
This work is supported by the B-type Strategic Priority Program of the Chinese Academy of Sciences, Grant No. XDB41000000, National Natural Science Foundation of China (Grants No. 12033010, 11573073, 11633009, 11773081, 11761131008), CAS Interdisciplinary Innovation Team, Youth Innovation Promotion Association and the Foundation of Minor Planets of Purple Mountain Observatory.

\end{CJK*}
\end{document}